\documentclass[aps,prl,superscriptaddress,notitlepage,amsmath,amssymb,nofootinbib,longbibliography]{revtex4-1}
\usepackage[utf8]{inputenc} 
\usepackage{graphicx}
\usepackage{url}
\usepackage[bookmarks, pagebackref=false]{hyperref}
\usepackage[usenames,dvipsnames]{xcolor}

\definecolor{bla}{HTML}{03396C}
\definecolor{blaa}{HTML}{005B96}
\definecolor{blaaa}{HTML}{6497B1}

\hypersetup{
  colorlinks, 
  bookmarksopen, 
  bookmarksnumbered,
  citecolor=blaa, 		
  linkcolor=blaa,    	
  urlcolor=blaa,			
}

\usepackage{natbib}

\newcommand{\nf}[1]{\ensuremath{\,\textcolor{blaa}{n}_{\textcolor{blaa}{f}}^{#1}}}
\newcommand{\MS}{\ensuremath{{\rm \overline{MS}}}}

\newcommand{\aMS}{\ensuremath{a_{\MS}}}
\newcommand{\aB}{\ensuremath{a_{\rm bare}}}
\newcommand{\aR}{\ensuremath{a_{\rm ren}}}
\newcommand{\aMOM}{\ensuremath{a_{R}}}

\newcommand{\aCCG}{\ensuremath{a_{\rm ccg}}}
\newcommand{\aQQG}{\ensuremath{a_{\rm qqg}}}
\newcommand{\aGGG}{\ensuremath{a_{\rm ggg}}}

\newcommand{\Zccg}{\ensuremath{Z_{\rm ccg}}}
\newcommand{\Zqqg}{\ensuremath{Z_{\rm qqg}}}
\newcommand{\Zggg}{\ensuremath{Z_{\rm ggg}}}

\newcommand{\Zc}{\ensuremath{Z_{\rm cc}}}
\newcommand{\Zq}{\ensuremath{Z_{\rm qq}}}
\newcommand{\Zg}{\ensuremath{Z_{\rm gg}}}

\newcommand{\bMS}{\ensuremath{\beta_{\MS}}}
\newcommand{\bMOM}{\ensuremath{\beta_{R}}}
\newcommand{\bCCG}{\ensuremath{\beta_{\rm ccg}}}
\newcommand{\bQQG}{\ensuremath{\beta_{\rm qqg}}}
\newcommand{\bGGG}{\ensuremath{\beta_{\rm ggg}}}
\newcommand{\bUNI}{\ensuremath{\beta_{\rm uni}}}
\newcommand{\asMZ}{\ensuremath{\alpha_s^{(5)}(M_Z)}}

\newcommand{\GammaB}{\ensuremath{\Gamma^V_{\rm bare}}}
\newcommand{\GammaR}{\ensuremath{\Gamma^V_{\rm ren}}}

\newcommand{\GamB}[1]{\ensuremath{\Gamma^{\rm #1}_{\rm bare}}}
\newcommand{\GamR}[1]{\ensuremath{\Gamma^{\rm #1}_{\rm ren}}}
\begin{document}

\title{
  \Large\color{bla} 
  Four-loop QCD MOM beta functions \\ from the three-loop vertices 
  at the
  symmetric point
}

\author{Alexander {\sc Bednyakov}}\email{bednya@theor.jinr.ru} 
\affiliation{Bogoliubov Laboratory of Theoretical Physics, Joint Institute for Nuclear Research,
Joliot-Curie 6, Dubna 141980, Russia}
\affiliation{P.N. Lebedev Physical Institute of the Russian Academy of Sciences, Leninskii pr., 5, Moscow 119991, Russia}
\author{Andrey {\sc Pikelner}}\email{pikelner@theor.jinr.ru} 
\affiliation{Bogoliubov Laboratory of Theoretical Physics, Joint Institute for Nuclear Research,
Joliot-Curie 6, Dubna 141980, Russia}

\begin{abstract}
	For the first time, we compute three-loop contributions to all triple vertices
	in QCD at the symmetric point. 
	The analytic results are obtained in massless QCD with an arbitrary color group in the Landau gauge.
	All new loop integrals are expressed in terms of harmonic polylogarithms at the sixth root of unity. 
	These corrections allow us to derive expressions for the four-loop QCD beta function 
	in a set of momentum-subtraction schemes.

\end{abstract}
\maketitle

\section{Introduction} 

The strong coupling constant $\alpha_s \equiv g_s^2/(4 \pi)$ is a fundamental
parameter of QCD. It enters the predictions of many Standard Model (SM)
observables, e.g., the Higgs production cross-section \cite{Anastasiou:2015ema} at the LHC,
 and 
the uncertainty of  
$\alpha_s$ significantly influences the comparison between theory and experiment. 
In itself,  $\alpha_s$ is not a physical observable, 
its value depends on the scale $Q^2$ and the utilized renormalization prescription (or scheme).
The freedom in choosing a normalization point 
manifests itself in ``running'' governed by the renormalization group (RG) equations 
\begin{align}
  \frac{d \alpha_s(Q^2)}{d \ln Q^2} = \beta ( \alpha_s(Q^2) ). 
  \label{eq:as_rge_general}
\end{align}
Given the renormalization scheme (RS), one can  calculate the $\beta$-function
order by order in perturbation theory (PT).
To parametrize the strength of strong interactions,  
one usually uses the \MS scheme and quotes the value of $\asMZ = 0.1179(10)$ \cite{Tanabashi:2018oca} defined 
in effective QCD with $n_f=5$ massless flavors and evaluated at the $Z$-boson mass.
This definition is of great convenience when studying inclusive  
observables dominated by short-distance effects. 
For almost 20 years the four-loop
$\beta^\MS$~\cite{vanRitbergen:1997va,Czakon:2004bu} in the $\MS$ scheme has been utilized in state-of-the-art QCD calculations, and only recently the five-loop result \cite{Baikov:2016tgj,Luthe:2017ttg,Herzog:2017ohr,Chetyrkin:2017bjc} became available in the literature. The latter 
allows one to significantly improve the agreement between $\alpha_s$ determinations (see, e.g., 
Ref.~\cite{Tanabashi:2018oca} for details) in precision measurements carried out  at different energy scales. 

At lower energies other $\alpha_s$ definitions can be more convenient. 
For example, in lattice QCD one can introduce the strong coupling as a particular RG-invariant combination 
(so-called \emph{invariant} charge) of the 
vertex and two-point Green functions evaluated at fixed external momenta in a fixed (usually Landau) gauge (see, e.g., Refs.~\cite{Boucaud:2013jwa,Zafeiropoulos:2019flq}).
Since 
lattice results \cite{Aoki:2019cca} 
give rise to 
one of the most precise determinations of \asMZ, it is important to study the strong coupling and its running in the momentum-subtraction (MOM) schemes. In these RSs the vertex functions and propagators are normalized in such a way that at a certain kinematic point 
there are no corrections beyond 
the tree level. 
As a consequence, 
the renormalized couplings coincide with the corresponding invariant charges and can be directly compared to the nonperturbative lattice results.

\begin{figure}[h]
  \centering
  \includegraphics{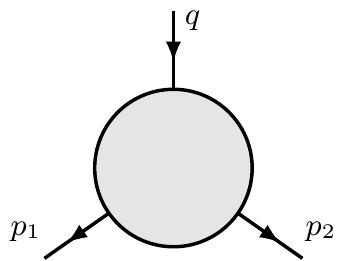}
  \caption{For symmetric point integrals $p_1^2 = p_2^2 = q^2$ and for auxiliary
    integrals we use $p_1^2 = p_2^2, q^2 = x\, p_1^2$}
  \label{fig:v3pt}
\end{figure}

Contrary to the $\MS$ scheme, in the MOM scheme 
one needs to know the Green 
functions beyond the divergent terms. 
Some choices of kinematics can make the calculation quite challenging
, especially of vertices. 
In this paper, we consider three-point functions that depend on momenta $p_1$, $p_2$, and $q=p_1+p_2$ (see Fig.~\ref{fig:v3pt}) and utilize the symmetric normalization point (SMOM) in the Euclidean region $p_1^2 = p_2^2 = q^2 = -Q^2$.
One-loop QCD renormalization in the SMOM scheme in a general linear gauge has
been known for quite a long time, following the pioneering work in Ref.\cite{Celmaster:1979km}. 
At the moment, only two-loop expressions for the QCD vertices in this kinematics
are available in the literature in numerical \cite{Chetyrkin:2000fd} and
analytical \cite{Gracey:2011vw} forms, allowing one to find three-loop
\cite{Chetyrkin:2000fd,Gracey:2011pf} strong-coupling beta functions in SMOM RS. 
In our work, we exploit modern Feynman-integral evaluation methods 
to improve these results by one more order of PT. 
To simplify our calculation, we routinely use the Landau gauge, in which the gluon propagator is transversal and there is no need to consider the gauge-parameter renormalization.

In the context of perturbative QCD (pQCD) one can relate renormalized couplings $\aR \equiv \alpha^{\rm ren}_s/(4 \pi)$ defined in different RSs. 
In what follows, we consider $\aR = \{\aGGG, \aCCG, \aQQG\}$, which satisfy  
\begin{align}
	\mu^{-2 \varepsilon} \aB & = Z_{\aR} \aR  
	=  \left[\frac{\Zggg^2}{\Zg^{3}}\right] \aGGG 
	=  \left[\frac{\Zccg^2}{\Zc^2 \Zg}\right] \aCCG
	=  \left[\frac{\Zqqg^2}{\Zq^2 \Zg}\right] \aQQG,
	\label{eq:bare_to_ren}
\end{align}
with $\aB$ being the bare coupling defined in the dimensionally regularized theory with $d=4-2\varepsilon$. The bare coupling is divergent and is related to the renormalized ones $\aR$ via divergent factors $Z_{\aR}$. The latter are combinations of the renormalization constants of the 
three-gluon ($\Zggg$), ghost-gluon ($\Zccg$), and quark-gluon ($\Zqqg$) vertices together 
with those of the gluon ($\Zg$), ghost ($\Zc$) and quark ($\Zq$) fields. 

The vertex renormalization constants denoted collectively by $Z_V$ are obtained order by order in perturbation theory from the bare $\GammaB$ vertices by making sure that the renormalized counterpart \GammaR
\begin{align}
	\GammaR(q_i, \aR) =Z_{V}(\aR) \GammaB(q_i, \aB), \quad \mu^{-2 \varepsilon} \aB = Z_{\aR} \aR
  \label{eq:mulren}
\end{align}
is finite for all external momenta $q_i$ and satisfies certain normalization conditions. 

In the \MS scheme only divergent terms are subtracted, and one can prove that $\aMS = \aGGG^\MS = \aCCG^\MS = \aQQG^\MS$. 
On the contrary,  in the SMOM schemes one also subtracts finite terms and requires that there are no $\mathcal{O}(\aR)$ corrections to a particular $\GammaR$ at the symmetric point,
characterized by momentum $Q^2$. 
To avoid the appearance of $\log (Q^2/\mu^2)$ in the renormalization constants, it is convenient to choose $Q^2 = \mu^2$. 

The vertex functions of our interest have color and space-time indices and one can decompose them in terms of basis tensors with Lorentz-invariant coefficients (form-factors). 
The choice of the basis is not unique and we make use of a
decomposition~\cite{Gracey:2011vw} taht is valid for the symmetric point. 
To save space, we present here only terms \cite{Chetyrkin:2000fd} that are relevant for the definition of the required vertex renormalization constants \eqref{eq:mulren}: 
\begin{align}
	\Gamma^{abc}_{\mu}(p_1,p_2) & = - i g_s f^{abc} \left( p^\nu_1 g_{\nu\mu} \Gamma^{\rm ccg}(-\mu^2) + \ldots \right),
\label{eq:v_ccg}\\
	\Gamma^{abc}_{\mu\nu\rho}(p_1,p_2) & =   i g_s f^{abc} \left( T_{\mu\nu\rho} \Gamma^{\rm ggg}(-\mu^2) + \ldots\right), 
\label{eq:v_ggg}\\
	\Gamma^{a}_{\mu,ij}(p_1,p_2) & = g_s T^a_{ij} \left( \gamma_\mu \Gamma^{\rm qqg}(-\mu^2) + \ldots\right). 
\label{eq:v_qqg}
\end{align}
Here all momenta are assumed to be outgoing, $p_1$ in Eq.\eqref{eq:v_ccg} corresponds to the antighost, and $T_{\mu\nu\rho} = g_{\mu\nu} (p_1 - p_2)_\rho + \ldots$ represents the tensor that enters into the tree-level three-gluon vertex. 
The SU(N) generators $T^a_{ij}$ in the quark-gluon vertex \eqref{eq:v_qqg} satisfy $[T^a, T^b]_{ij} = i f^{abc} T^c_{ij}$ with structure constants $f^{abc}$. 

The expressions for the bare form factors $\GammaB$ with $V = \{ {\rm
  ggg,~ccg,~qqg} \}$ are extracted from the tensor vertices by means of the projectors given in Ref.~\cite{Gracey:2011vw}.  
To define $\aGGG$ in the respective SMOM scheme (MOMggg) via Eq.~\eqref{eq:bare_to_ren}, we require that at the symmetric point $\GamR{ggg}(-\mu^2) = 1$, i.e. $\Zggg^{-1} = \GamB{ggg}(-\mu^2)$. 
In the same way one can relate the bare coupling $\aB$ to $\aCCG$ or $\aQQG$ by requiring that either $\GamR{ccg}(-\mu^2) = 1$ (MOMh), or $\GamR{qqg}(-\mu^2) = 1$ (MOMq).
Since field renormalization constants also enter into Eq.\eqref{eq:bare_to_ren}, we have to impose conditions on $\Zg$, $\Zc$, and $\Zq$ in the SMOM scheme: there should be no corrections to the corresponding tree-level propagator for the external momentum $q^2=-\mu^2$.

From Eq.~\eqref{eq:bare_to_ren} one can deduce that two renormalization prescriptions for \aR, say $\aMOM$ and $\aMS$, are related via finite correction factors $X_R$:
\begin{align}
	\aMOM & = \left( Z_{\aMOM} / Z_{\aMS} \right) \aMS \equiv \aMS X_R = \aMS 
	\left[1 + \sum_l X^{(l)}_R \aMS^l \right].
	\label{eq:RS_change}
\end{align}
Given $X_R$ at $L$ loops, one can determine the  $(L+1)$ MOM-scheme beta
functions from $\bMS$ via a relation that is valid in the Landau gauge,
\begin{align}
	\bMOM & \equiv \frac{ d \aMOM}{d \ln \mu^2}  =  
	\frac{\partial \aMOM (\aMS)}{\partial \aMS} \cdot \bMS(\aMS), \qquad \aMS = \aMS(\aMOM),
	\label{eq:bMOM_bMS}
\end{align}
where in the final step we invert Eq.~\eqref{eq:RS_change} to express $\bMOM$ in terms of $\aMOM$. 
The main aim of this paper is to calculate the three-loop corrections
$X^{(3)}_R$ to the relations \eqref{eq:RS_change} between $\aGGG$, $\aCCG$, $\aQQG$ and $\aMS$. 
As one of the applications of our result, we use  the four-loop beta function in
the \MS scheme \cite{vanRitbergen:1997va,Czakon:2004bu} 
to find the corresponding beta functions $\bGGG$, $\bCCG$, and $\bQQG$ in the
considered SMOM schemes.

\section{Details of the calculation}
\label{sec:calc}

We generate Feynman diagrams 
with \texttt{DIANA}~\cite{Tentyukov:1999is},  and obtain 
8, 106, and 2382 graphs for the 
$\GamR{ggg}$ 
three-gluon vertex at one, two, and three loops, respectively.
Both 
ghost-gluon
and 
quark-gluon
vertices give rise to 2, 33, and 688 diagrams at the same loop levels. 
After the application of projectors~\cite{Gracey:2011vw} and taking fermion and
color traces\cite{vanRitbergen:1998pn}, we are left with scalar Feynman integrals,
which we reduce to a set of two one-loop, eight two-loop, and 51 three-loop master integrals by means of  
\texttt{Reduze 2}\cite{vonManteuffel:2012np} and \texttt{FIRE6}\cite{Smirnov:2019qkx}. 

The main challenge is calculating of the full set of three-loop three-point integrals in SMOM kinematics, which was not available in the literature. 
To compute master integrals, we rely on the \emph{linear reducibility} of
massless vertex-type integrals with arbitrary off-shell momenta, which is 
proven to take place up to three-loop order\cite{Chavez:2012kn,Panzer:2014gra}. 
Due to the latter  property, for our integrals in 
more restricted kinematics 
we choose two
strategies of evaluation. 
First of all, we 
derive 
a new basis of finite master integrals 
\cite{vonManteuffel:2014qoa} 
and try to compute them directly by means of the \texttt{HyperInt} package\cite{Panzer:2014caa} in terms of generalized polylogarithms (GPLs).
Unfortunately, we are unable to calculate the most complicated integrals and 
instead use another
strategy based on the solution of the system of differential equations (DEs).
The method of DEs cannot be applied directly to the calculation of single scale integrals
we are interested in, and we construct a set of auxiliary integrals with arbitrary
external $q^2 = x\,p_1^2$ (Fig.\ref{fig:v3pt}). After switching to a new variable $z$,
$x = 2-z-1/z$,  
we reduce the original DE system 
to the so called
$\varepsilon$-form\cite{Henn:2013pwa}. 
Due to the presence of singularities at complex points, we make use of the \texttt{epsilon}
package\cite{Prausa:2017ltv}, which is capable of dealing with the latter. 
The obtained system in the $\varepsilon$-form 
is 
easily solved
order by order in the $\varepsilon$ expansion.
The solution is given by 
linear combinations of 
GPLs $G(a_1,\dots,a_n;z)$, where $a_i$ correspond to different sixth roots of 
unity, 
with a number of unknown constants to be fixed from boundary conditions. 
The latter can be 
obtained  
by matching the expansion in the limit $q^2 \to 0$, corresponding to 
$z \to 1$, with the explicit result of 
large-momentum asymptotic expansion. 
Since $z=1$ is a singular point of the DEs, naive Taylor expansion is not sufficient
to fix all of the constants
and we exploit 
the \texttt{EXP} package\cite{Harlander:1997zb,Seidensticker:1999bb} to generate the series in terms of massless propagators. We compute the latter by the
\texttt{MINCER} package \cite{Gorishnii:1989gt,Larin:1991fz} keeping exact dependence\footnote{Available at
  \url{https://www.nikhef.nl/~form/maindir/packages/mincer/mincerex.tgz}.}
on the space-time dimension variable.

  In this way, we obtain the analytic results for auxiliary integrals depending on $z$. 
  Taking the limit $z\to e^{i \pi/3}$, which is regular and corresponds to SMOM kinematics,
  we 
  compute the required single-scale master integrals. 
  The correctness of the analytic calculation is verified numerically using \texttt{pySecDec}\cite{Borowka:2017idc}.
  Substituting the integrals in the expressions for the three-loop $\GammaB$ and expanding in $\varepsilon$ up to the necessary order, we see that the maximal transcendental weight 
  in the final result is, as expected, $2L = 6$ with $L$ being the loop order.
Using the basis and the reduction rules from Ref.\cite{Henn:2015sem},
we are able to simplify the expressions for $\GammaB$ significantly. 
As a consequence, we obtain the SMOM renormalization constants, conversion
factors \eqref{eq:RS_change}, and the beta functions \eqref{eq:bMOM_bMS} in a
rather compact form.

\section{Results and conclusion}

\label{sec:results}

All of the necessary renormalization constants up to three-loop order were calculated iteratively via Eq.~\eqref{eq:mulren}. 
We also reproduced the well-known three-loop $\MS$ expressions, required to   
derive the relations \eqref{eq:RS_change} between the SMOM couplings and $\aMS$ evaluated at the same scale $Q^2 = \mu^2$. 
To save space, we present the results in numerical form with all QCD color
factors substituted explicitly:
\begin{align}
  X_{\rm ccg} = 1 & + \aMS \left(18.54827536 - 1.111111111 \nf{}\right) \nonumber\\
                  & + \aMS^2 \left(641.9400677 - 85.55595017 \nf{} + 1.234567901 \nf{2}\right) \nonumber\\
                  & + \aMS^3 \left(26810.13185 - 5350.674817 \nf{}+ 240.8472277 \nf{2} - 1.371742112 \nf{3}\right),\label{eq:XCCG}\\
  X_{\rm qqg} = 1 & + \aMS \left(16.71577458 - 1.111111111 \nf{}\right) \nonumber\\
                  & + \aMS^2 \left(472.1590958 - 83.11121681 \nf{}+ 1.234567901 \nf{2}\right) \nonumber\\
                  & + \aMS^3 \left(16997.21982 - 4340.986026 \nf{}+ 228.6939963 \nf{2} - 1.371742112 \nf{3}\right),\label{eq:XQQG}\\
  X_{\rm ggg} = 1 & + \aMS \left(26.49248887 - 3.416806434 \nf{}\right) \nonumber\\
                  & + \aMS^2 \left(960.4627178 - 202.0850109 \nf{}+ 7.687393017 \nf{2}\right) \nonumber\\
                  & + \aMS^3 \left(42285.00716 - 12133.42891 \nf{}+ 902.7134506 \nf{2} - 14.34154686 \nf{3}\right)\label{eq:XGGG}.
\end{align}

The analytic results for the general gauge group are available as a Supplementary Material and have several remarkable properties. 
We managed to simplify the two-loop part \cite{Gracey:2011pf} and expressed it
solely in terms of powers of $\pi$, odd $\zeta$ values, and two polygamma
functions $\psi^{(1)}(1/3)$ and $\psi^{(3)}(1/3)$.
It turns out that to write down the three-loop contribution we only need to introduce three additional constants: $\psi^{(5)}(1/3)$,
and two combinations $H_5$ and $H_6$ of GPLs with uniform transcendental weights five and six, respectively.  
Evaluating $H_{5,6}$ numerically \cite{Vollinga:2004sn} with high precision and using the \texttt{PSLQ} algorithm\cite{ferguson1999analysis},
we 
reconstructed 
the new constants\footnote{A similar set of constants generated by cyclotomic
  polylogarithms was considered in Ref.\cite{Ablinger:2011te}.} through a more restricted basis \cite{Kniehl:2017ikj} of real
parts of harmonic polylogarithms of the argument $e^{i\pi/3}$.

As a first application of Eqs.\eqref{eq:XCCG},\eqref{eq:XQQG}, and \eqref{eq:XGGG} we obtain a set of SMOM beta functions
via Eq.\eqref{eq:bMOM_bMS}, thus extending the results of
Ref.~\cite{Gracey:2011pf} to four-loop order:
\begin{align}
  \bCCG & = \bUNI(\aCCG) - \aCCG^4 \left(2813.492952 - 617.6471546 \nf{}+ 21.50281811 \nf{2}\right) \nonumber\\
        & - \aCCG^5 \left(96089.34786 - 23459.32128 \nf{}+ 1735.992218 \nf{2} - 33.24145137 \nf{3}\right),\label{eq:bCCG}\\
  \bQQG & = \bUNI(\aQQG) - \aQQG^4 \left(1843.652731 - 588.6548459 \nf{}+ 22.58781183 \nf{2}\right) \nonumber\\
        & - \aQQG^5 \left(68529.68547 - 15466.43194 \nf{}+ 1093.568841 \nf{2} - 18.85323795 \nf{3}\right),\label{eq:bQQG}\\
  \bGGG & = \bUNI(\aGGG) - \aGGG^4 \left(1570.9844 + 0.56592607 \nf{}- 67.089536 \nf{2} + 2.6581155 \nf{3}\right) \nonumber\\
        & - \aGGG^5 \left(94167.261 - 27452.645 \nf{}+ 4152.5388 \nf{2} - 543.68484 \nf{3} + 20.429348 \nf{4}\right),\label{eq:bGGG}
\end{align}
with $\beta_{\rm uni}(a) = - a^2 (11 - 2/3 \nf{}) - a^3(102 - 38/3\nf{})$ being the universal two-loop scheme-independent contribution.

The obtained formulas can be used in a number of ways. 
For example, one can improve the precision of matching between the lattice and pQCD results.
In addition, the possibility to switch from one RS to another in truncated PT
series for observables, or more generally RG-invariant quantities (see, e.g.~Refs.~\cite{Gracey:2014pba,Herzog:2017dtz}), provides us with an additional handle on 
theoretical uncertainties beyond simple scale variation. 
Moreover, the computed 
integrals can also be used in SMOM calculations of more complicated three-point
Green functions with operator insertions, e.g., for studies of the light-quark masses as in Refs.~\cite{Almeida:2010ns,Gracey:2011fb}.

\acknowledgments

We thank A. Kotikov, V. Magerya, and S. Mikhailov for fruitfull discussions.
The work of A.P. is supported by the Foundation for the Advancement
of Theoretical Physics and Mathematics ``BASIS.''
The work of A.B. is supported by the Grant of the Russian Federation Government, Agreement No. 14.W03.31.0026 from 15.02.2018.
\bibliography{bSMOM4l}
\end{document}